
\font\titlefont = cmr10 scaled \magstep2
\magnification=\magstep1
\vsize=22truecm
\voffset=1.75truecm
\hsize=15truecm
\hoffset=0.95truecm
\baselineskip=20pt

\settabs 18 \columns

\def\b{\bigskip}
\def\bb{\bigskip\bigskip}

\def\ce{\centerline}

\def\no{\noindent}

\rightline{AMES-HET-94-07}
\rightline{June 1994}
\rightline{(Revised)}
\bb

\ce{\titlefont {Electroweak Sphaleron For Effective Theory}}
\ce{\titlefont {In The Limit Of Large Higgs Boson Mass}}

\b

\ce{ X. Zhang, ~~ B.-L. Young and S.K. Lee}
\b
\ce{Department of Physics \& Astronomy }
\ce{ Iowa State University,}
\ce{ Ames, Iowa
 50011}

\b

\bb
\ce{\bf ABSTRACT}
\b
 Theoretical arguments suggest that the Higgs sector of the standard model
is an effective theory.
We parametrize the new physics by means of an effective
 Lagrangian technique and
 study its effect on the energy of the electroweak
sphaleron. We found that
in the presence of a certain class of
higher dimension operators
the sphaleron energy
becomes arbitrarily large as the Higgs boson mass, $m_H$, increases.
 The physical meaning of this result and its implications
 to electroweak
baryogenesis in the
dynamical symmetry breaking models are discussed.
\bb

\bb
\filbreak

 The energy of the sphaleron in the one-doublet Higgs model,
 calculated first by
Manton and Klinkhamer
[1], then improved in [2],
 is given by $E^{sphal} = {2 M_W \over \alpha_W}
         B( \lambda /  g^2 )$, where
$M_W$ is the W-boson mass,
$\alpha_W = {g^2 / 4 \pi} $ and
$g$ is the SU(2) coupling constant; $B$
is a function of the Higgs boson mass, $m_H^2 = 2 \lambda v^2$
with $v$ being the Higgs
vacuum expectation value.
Thus the $E^{sphal}$ will be fixed once
the Higgs boson mass is known.
Experimentally,
LEP data put a lower limit on $m_H$, which is about 60 GeV[3].
 There are also theoretical arguments
which give an upper bound on $m_H$. For instance, triviality arguments
suggest that $m_H < 600-800$ GeV[4]. For this range of $m_H$, the
  factor $B( {\lambda \over
g^2 })$ does not change very much. Actually,
  the numerical evaluation gives
$1.5 \leq B \leq 2.7$[1, 2] when $m_H$
 varies from zero to infinity[5]. Thus
  the weakly interacting theory
with a light Higgs boson is indistinguishable from
 the strongly coupled theory in the infinite Higgs boson
mass limit as far as the $E^{sphal}$ is concerned.

There are the theoretical arguments of ``Triviality"[4] and ``Naturalness"[7]
against the
elementary
scalar sector of the standard model (SM), and one believes
that the Higgs sector of the standard model is an effective theory.
In this paper
we will
study the properties of the sphaleron solution in an effective theory and
demonstrate that
$E^{sphal}$ will
 increase without bound as the Higgs boson mass goes to infinity.

First of all,
we consider the effective lagrangian (EL) with a linear realization of
the SM gauge symmetry, where
  the effects of the new physics are parametrized by
a set of higher dimension operators
in addition to those present in the SM.
Within such an EL,
 we will look for the sphaleron solution and calculate its energy.
 In Ref[8], it was shown that the dimension
6 operators have a very small effect on the $E^{sphal}$. However, starting at
dimension 8, there are operators,
which can make $E^{sphal}$ diverge in the heavy
Higgs boson mass limit. A dimension 8 operator in question is

$${
  {\cal O} \sim {1 \over \Lambda^4 }
            {  \{ {(D_\mu \Phi)}^\dagger D^\mu \Phi \} }^2 ~~~. }\eqno(1)$$

\no There are other operators with dimension greater than 8, which may
also give
infinite
contribution
to the $E^{sphal}$. However,
we will concentrate here on the operator $\cal O$ for detail discussion.

First of all, let us write down explicitly the EL relevant to
our discussions.

$${
\eqalign{
{\cal L}^{eff} = & -{1\over 4} F_{\mu\nu}^a F^{a \mu\nu} -{1\over 4}
                  f_{\mu\nu} f^{\mu\nu} + {(D_\mu \Phi)}^\dagger D^\mu \Phi
                    - V( \Phi ) \cr
                & + {1 \over \Lambda^4 } {  \{ ( D_\mu \Phi )^\dagger
                 D^\mu \Phi  \} }^2              ,\cr }
}  \eqno(2.a)$$

\no where $F_{\mu \nu}^a$ are the
$SU_L(2)$ field strength,
$f_{\mu\nu}$ the
$U_Y(1)$ field strength, and

$${
V( \Phi ) = \lambda ( \Phi^\dagger \Phi - {v^2 \over 2} )^2;
{}~~ D_\mu \Phi = \partial_\mu \Phi -i{g\over 2}
 {\vec \tau} \cdot {\vec W}_\mu \Phi -i {g^\prime \over 2}
             {\vec B}_\mu \Phi ~, } \eqno(2.b)$$
\no where $W_\mu$ and
$B_\mu$ are the gauge fields of $SU_L(2)$
and $U_Y(1)$
respectively. In (2.a), we have not included the fermion fields since we
consider
 the sphaleron solution in the bosonic sector of the effective theory.

Klinkhamer and Manton[1] have shown that in the SM,
 there is a saddle-point
 solution, which is the sphaleron. Since the higher dimension operator
$\cal O$ does not affect the symmetry breaking pattern of the
minimal model, nor
change the topology of the field configuration space,
the sphaleron solution should exist in the ${\cal L}^{eff}$ (2.a).

Following the usual procedure
we consider the static solution by setting the time components
of the gauge fields to zero. Also, as is generally done,
 we will set the Weinberg angle to zero, so that the $U_Y(1)$
gauge field is decoupled and may be consistently set to zero.
 Following Ref.[1],
 we use the spherical symmetric ansatz for the gauge field
$\vec {W_\mu}$ and
the Higgs field $\Phi$,

 $${
W_i^a \sigma^a dx^i = - {2 i\over g} f( \xi ) ~dU^{\infty} {( U^{\infty} )}
                ^{-1} ~~~; } \eqno(3.a)$$

$${
\Phi = {v \over {\sqrt2}} ~ h( \xi ) ~ U^\infty { \pmatrix{0 \cr
                                1 \cr} } ~~~~~, }\eqno(3.b)$$

\no where
$\xi = g v r$ and,
$${
U^\infty = {1 \over r} \pmatrix{ z & x + iy \cr
                              -x + iy & z \cr } ~~~. }$$

The energy functional is given by

$${ \eqalign{ E^{sphal} = &
           {4 \pi v \over g} \int_0^{\infty} d \xi \{ 4 ({df \over d\xi })^2
                         + {8 \over \xi^2 } f^2 (1-f)^2  + {1\over 2}
                \xi^2 ({ d h \over d\xi })^2 \cr
                & + h^2 (1-f)^2 + {1\over 4} {\lambda \over g^2 }
                    \xi^2 (h^2 - 1)^2 \cr
             & + c ~[ \xi^2 ({dh \over d\xi })^4 + 4 h^2 ({dh \over d\xi})^2
                            (1-f)^2 + 4 {h^4 \over \xi^2 }(1-f)^4 ~]~~, \cr }
{}~~~~~} \eqno(4)$$

\no where $c \equiv ({g^2 \over 4}) {v^4 \over \Lambda^4 }$.
It is straightforward to derive the Euler-Lagrange equation for
this energy functional. They read,

$${
\eqalign{ \xi^2 {d^2 f \over d\xi^2 }
          = & 2 f (1-f) (1-2f )- {\xi^2 \over 4} h^2 (1-f) \cr
           & - ~c ~ \{ \xi^2 h^4 ({dh \over d\xi })^2 (1-f )
            + 2 h^4 (1-f)^3 \} ~~; \cr }
} \eqno(5.a)$$

$${
\eqalign{ {d \over d\xi }( \xi^2 {dh \over d \xi} ) = &
                       2 h (1-f)^2 + ({ \lambda \over g^2 })
                            \xi^2 (h^2 - 1) h \cr
                   & + ~8  c~ [ h ({dh \over d\xi })^2 (1-f)^2
                  + 2{h^3 \over \xi^2} (1-f)^4 ] \cr
                   & - ~4 c~[ {d \over d\xi}( \xi^2 ({ dh \over d \xi })^3 )
                     + 2 {d \over d \xi }( {dh \over d\xi } h^2 (1-f)^2 )
                         ] ~~ . \cr}
} \eqno(5.b)$$

\b

\no The boundary conditions for $f( \xi )$ and
$h( \xi )$ are given by

$${
f( \xi ) \rightarrow \xi^2 ~~{\rm and} ~~h( \xi ) \rightarrow
                  \xi ~~~{\rm for} ~~ \xi \rightarrow 0 ~;}\eqno(6.a)$$

$${
f( \xi ) ~ ~{\rm and} ~~h( \xi ) \rightarrow 1 ~~{\rm for} ~~\xi
    \rightarrow \infty . } \eqno(6.b)$$

\no Note that these are
the same boundary conditions
as those[1] in the absence
of the higher dimension operator $\cal O$.
We numerically
integrate Eq.(4) by minimizing $E^{sphal}$
 for a given value of the
parameter $\lambda \over g^2$ in the range of zero to $10^{10}$[9].
In Fig.1 we plot
the $E^{sphal}$ of the present case together with that
of the standard model.
 One can see that for small $\lambda \over g^2$, the new
 physics effect on $E^{sphal}$ is negligible
and the two energies are
practically indistinguishable. However, it becomes important
for larger $\lambda \over g^2$. For instance, for ${\lambda \over g^2}
> 10^5$,
the sphaleron energy has exceeded the maximal value of the SM sphaleron.
{}From our numerical results we can extract
  the behavior
of the sphaleron energy in the large Higgs boson mass limit.
  Approximately we have
 $E^{sphal} \sim
 ({ \lambda
\over g^2 })^{1\over 4}$ for
${\lambda \over g^2} > 10^5$ . This behavior can also be understood by
using a particularly simple {\it ansatz},
 considered by Manton and Klinkhamer[1],
for the radial functions $f (\xi )$ and
$h ( \xi )$,

$${
   f( \xi ) = \cases{ $( ${ \xi \over \Xi}$ )^2$, & $ \xi \leq \Xi $ \cr
                        1, & $\xi \geq \Xi$ , \cr }
}   \eqno(7.a)$$

$${
  h( \xi ) = \cases{ $ \xi \over \Omega$ , & $\xi \leq \Omega $ \cr
                    1, & $\xi \geq \Omega$ , \cr }
}  \eqno(7.b)$$

\no where $\Omega$ and
$\Xi$ are two variational scale parameters.
With such an {\it ansatz}, the dominant terms in $E^{sphal}$
 for very large $\lambda \over g^2$ are given by

$${
E^{sphal} \sim {4 \pi v \over g} {1\over 210}
\{ 4 ~({\lambda \over g^2} )~
 \Omega^3 ~ + ~ 210~ c~ {97 \over 15} {1\over \Omega}
     \}  . }  \eqno(8)$$

\no The scale parameter $\Omega$ that minimize (8) goes like
$({ \lambda \over g^2 })^{- 1/4}$, which gives
$${
E^{sphal} \sim ({ \lambda \over g^2 } )^{1\over 4} ~~~.
 } \eqno(9)$$


One remark here is that
 the blowup of the sphaleron energy
is caused technical by the ``frozen" Higgs field, $h( \xi ) \equiv 1$.
To see it clearly, let us rewrite the $E^{sphal}$ in terms of $\Omega$.
 We have
$E^{sphal} \sim {1 / \Omega}$.
 From Eq.(7.b), $\Omega \rightarrow 0$ means that
$h( \xi )$ \rightarrow 1$.
Because of the triviality of the
Higgs sector, however,
the physical meaning of the
$m_H \rightarrow \infty$ limit is questionable. In the following
 we will argue that our results indicate that the $E^{sphal}$
is infinity
 in the non-linear EL of the
dynamical symmetry breaking (DSB) models.

To proceed with the discussion of the sphaleron in the effective theory
of a DSB model,
 let us consider a limit
$\lambda \rightarrow \infty$
in ${\cal L}_{eff}$. In this limit, the Higgs field $\Phi$
  can be parametrized as
$\Phi = {v \over {\sqrt 2}} \Sigma \pmatrix{0 \cr 1 \cr}$, where
  $\Sigma$ is a $2 \times 2$ matrix for the Goldstone boson fields.
Thus ${\cal L}^{eff}$ in (2.a) becomes

$${
{\cal L}_\sigma = - {1 \over 4}F^a_{\mu \nu} F^{a \mu \nu}
 + {v^2 \over 4}TrD_\mu \Sigma^\dagger D^\mu \Sigma ~+ ~
              {v^4 \over{ 4^2 \Lambda^4} } ( TrD_\mu \Sigma^\dagger
                         D^\mu \Sigma )^2
, }\eqno(11)$$

\no where we have neglected the $U_Y(1)$ field.
The ${\cal L}_\sigma$ has the form of the gauged non-linear sigma models
and will
 describe the low energy physics of dynamical electroweak
symmetry breaking
 models.

 The sphaleron solution
of ${\cal L}_\sigma$[10] has been considered by Klinkhamer and Boguta[11].
   They use the following
spherical symmetric ansatz, which is a gauge transformation
of that in eq.(3),

$${
\Sigma = 1 ; ~~~~
{g \over 2} W_i = { 1 - f \over r^2 }~ {({\vec r} \times {\vec \tau})}_i
{}~, }\eqno(12)$$

\no with the boundary conditions $f(0) = 0 ~{\rm and} ~ f( \infty ) = 1 $.
The sphaleron energy functional is

$${
E^{sphal}_\sigma = {4 \pi v \over g} \int d \xi \{
          4 ({df \over {d\xi}})^2 + {8 \over \xi^2} f^2 {(1-f)}^2
+{(1-f)}^2 + ( 4 c ){ {(1-f)}^4 \over \xi^2 }
       \} ,  }   \eqno(13)$$

\no where $c = {g^2 \over 4} {v^4 \over \Lambda^4}$.
Klinkhamer and Boguta noticed that
 the $E^{sphal}_\sigma$ blows up[12] because of the last term.
Actually,
 the sphaleron energy density is singular and the boundary condition
is not satisfied by the differential equation for $f( \xi )$,
which minimizes $E^{sphal}_\sigma$.

We realize that the singularity in
$E^{sphal}_\sigma$
is removable once the sphaleron solution
of ${\cal L}_\sigma$ is understood as the
sphaleron solution of ${\cal L}^{eff}$ (2.a)
in the limit  ${\lambda \over g^2}
\rightarrow \infty$.
  In fact, in the limit
${\lambda \over g^2 } \rightarrow \infty$, $h( \xi ) \rightarrow 1$,
then
${\cal L}^{eff}$ goes to
${\cal L}_\sigma$ and
 $E^{sphal} \rightarrow E^{sphal}_\sigma$.
Thus in the non-linear EL ${\cal L}_\sigma$, the sphaleron energy is
indeed divergent.
 To illustrate the physical meaning of this result,
we consider a toy DSB model.
This is the
 one family standard model, but no elementary Higgs field is introduced.
In this model the SM gauge symmetry is broken by the quark condensate
driven by the QCD interaction, where both the electroweak symmetry breaking
scale and the weak gauge boson masses are, of course, very low.

It is well-known that the strong interaction of a DSB model can
be described by an EL similar to ${\cal L}_\sigma$,
 when the heavy fermions, i.e., the
quarks, are frozen out at
an energy below the DSB strong interaction scale.
Below the DSB strong interaction scale, there are only leptons in the fermion
sector,
and the lepton
number current is violated by an SU(2) anomaly which involves two
SU(2) currents. However, its
amplitude will vanish according to our results since
the sphaleron energy
is infinity.

 Is the above reasonable and what is its significance? In the
fundamental Lagrangian of quarks and leptons,
there are two kind of fermion number currents, one lepton number,
and the other baryon number. Both have an
SU(2) anomaly, however, their difference is anomaly free,
which means that the total change
of the lepton number must equal to that of the baryon
 number.
Since baryon fields do not exist in the low energy Lagrangian
lepton number violation processes are forbidden. In other words,
 the sphaleron energy
should be infinity.

Our results have direct implications
 on the electroweak baryogenesis.
As is argued in Ref.[13], to avoid the washout of the baryon asymmetry,
the following condition is needed
$${
{E^{sphal} ( T_t ) \over T_t } > 45 ~~~~, } \eqno(14)$$

\no where $T_t$ denotes the electroweak
phase transition temperature.
In computing
$E^{sphal}(T_t)$, one should use the full effective action, which
includes terms
depending on derivatives, denoted by $D_T$, and terms independent of
derivatives, which are generally known as
the
  effective potential, denoted as $V_T$.
 The $V_T$
in the standard model and its extensions has been studied in detail
in recent years[14]. And the sphaleron solution with a temperature
dependent $V_T$ has also been studied by Braibant, Brihaye and Kunz in
 Ref.[15].
They concluded that for the temperature dependent $V_T$ in the
absence of a cubic term, the
 sphaleron energy as a function of
the temperature is given by
$${
E^{sphal}( T ) = E^{sphal}(T=0) ~~{ v( T ) \over v}
{}~~~~, } \eqno(15)$$
\no where $v( T )$ is the vacuum expectation value of the Higgs
field at the temperature $T$.
With a cubic term in the $V_T$, Eq.(15) has
to be modified, but it remains to be a rather good
approximation[15].
The finite temperature correction to $D_T$ in the standard model
has been considered by Dine, Huet and Singleton
in Ref.[16]. They estimated the contribution of several typical
Feynman diagrams and argued that the correction
to $E^{sphal}(T)$ is about $20 \%$.

In the effective theory we consider in this paper, the higher dimension
operators will contribute to $V_T$ as well as
$D_T$. As a result, the SM Higgs boson mass limit required by
electroweak baryogenesis will be changed.
Assuming the validity of Eq.(15),
 the higher dimension operators
can alter this upper limit for the Higgs boson mass in two ways; 1)
it changes the usual relation between $v( T )$ and the Higgs boson mass; 2)
it changes the dependence of the $E^{sphal}(T=0)$
 on the Higgs boson mass. For small
Higgs boson mass, the impact of the higher dimension operator on
$E^{sphal}(T=0)$ is negligible. However, the relation between
$v( T )$ and
$m_H$
is changed due to a dimension 6 operator in $V_T$[17].
As a result,
the Higgs mass limit can be relaxed to the
experimental allowed region.
For large Higgs boson mass, as discussed in this paper,
the presence of higher dimension
operators will change $E^{sphal}(T=0)$ significantly, and
will relax the Higgs mass limit
further. In particular, in the dynamical symmetry
breaking theory, if the physics in
the true vacuum can be described by an EL
 ${\cal L}_\sigma$
 with
temperature dependent $v( T )$, which is now the composite Goldstone
boson decay constant,
  it will help to avoid the washout of the baryon asymmetry
because of the infinity sphaleron energy[18].
Certainly, to
construct a successful
 DSB
model for the electroweak baryogenesis,
 one must examine in detail the phase transition
and the source of CP violation.

In summary,
 we have taken the standard model Higgs sector
as an effective theory, then studied its sphaleron solution
 and calculated the sphaleron energy
in the presence of a dimension 8 operator.
 We found that the
 sphaleron energy diverges in the large Higgs boson mass limit.
This implies that
the sphaleron in
  ${\cal L}_\sigma$ of the DSB models has an
 infinity energy and will be helpful in preventing washout of
 the baryon number asymmetry.

\b

\bb

We thank K. Whisnant for reading the manuscript.
This work is supported in part by the Office of High Energy
 and Nuclear Physics of the U.S. Department of Energy
 (Grant No. DE-FG02-94ER40817).
\b
\bb
\vfill\eject
\ce{\bf {Figure Caption}}
\no [Fig.1] ~~Sphaleron energy as function of $\lambda / g^2$. The vertical
axis is the sphaleron energy in units of TeV and
the horizontal axis is $\log_{10}( \lambda / g^2 )$. The solid curve is
for the present case and the dash curve for the standard model.
In our numerical calculation we take $\Lambda = 1$ TeV. A different value of
$\Lambda$ will change the absolute value of the $E^{sphal}$, but
will not modify the behavior of the $E^{sphal}$ in the large Higgs boson mass
limit.

\ce{\bf References}

\item{[1]}N.S. Manton, Phys. Rev. D28, 2019 (1983);
         F.R. Klinkhamer and N.S. Manton, Phys. Rev. D30, 2212 (1984).

\item{[2]}
      T. Akiba, H. Kikuchi and
T. Yanagida, Phys. Rev. D38, 1937 (1988); L.G. Yaffe, Phys. Rev. D40, 3463
 (1989); J. Kunz and Y. Brihaye, Phys. Lett. B216, 353 (1989).



\item{[3]}G. Coignet, Plenary talk at the XVI International
         Symposium on Lepton-Photon Interactions, Cornell Univ.,
           Ithaca, New York, 10-15 Aug. 1993.

\item{[4]}For reviews, see, D. Callaway, Phys. Rept. 167, 241 (1988);
          H. Neuberger, in the Proceedings of
the XXVI International Conference on HIGH ENERGY PHYSICS, Volume II, P1360,
(1992) ed. J. Sanford.

\item{[5]}{The stability of the sphaleron energy against the changes of the
model parameters in multi-Higgs models with or without Higgs singlets
  has also been demonstrated in Ref.[6].}

\item{[6]}B. Kastening, R.D. Peccei and X. Zhang, Phys. Lett. B266, 413 (1991);
 B. Kastening and X. Zhang, Phys. Rev. D45, 3884 (1992);
         K. Enqvist and I. Vilja, Phys. Lett. B287, 119 (1992).

\item{[7]}G. 't Hooft, in Recent Developments in Gauge Theories, eds.,
            G. 't Hooft (Plenum Press, New York, 1980).

\item{[8]}
S. Lee, J. Spence and B.-L. Young,
 IS-J 5110, June 1993.


\item{[9]}Numerical analysis apparently does not allow the investigation
of the ${\lambda \over g^2} \rightarrow \infty$ limit with arbitrarily
large $\lambda\over g^2$. For very large
$\lambda \over g^2$, the term in
$E^{sphal}$ which is proportional to $\lambda \over g^2$ dominates the
numerical integration and a numerical instability will
appear. In our case, we can fairly efficiently calculate $E^{sphal}$ and
the radial functions $f( \xi )$ and
$h( \xi )$ for
${\lambda \over g^2 } \leq 10^{10}$.

\item{[10]} Since this is an effective theory with cut-off
$\Lambda \sim O( {\rm TeV})$, which is less than
 the mass of the sphaleron,
 one may wonder about
the reliability of our results.
However, in determining the
region of the applicability of the effective theory, it is the size
(inverse momentum of the typical fluctuation), not
the mass of the sphaleron, that is relevant. The size of the sphaleron is the
order of $1/ M_W$. So our approach is
valid since $\Lambda >> M_W$.

\item{[11]}F. Klinkhamer and J. Boguta, Z. Phys. C40, 415 (1988).

\item{[12]}{Klinkhamer and Boguta[11]
  interpreted the result in a different way. Our
 physical conclusion is different from theirs.}

\item{[13]}M. Shaposhnikov, Nucl. Phys. B287, 757 (1987); B299 (1988) 797.

\item{[14]}P. Arnold and O. Espinosa, Phys. Rev. D47, 3546 (1993); J.R.
Espinosa, IEM-FT-85/94, May 1994; and references therein.

\item{[15]}S. Braibant, Y. Brihaye and J. Kunz, THU-93/01, January 1993.

\item{[16]}M. Dine, P. Huet and R. Singleton, Jr., Nucl. Phys. B375 (1992) 625;
 M. Dine, {\it et. al.}, Phys. Rev. D46,
 550 (1992).

\item{[17]}X. Zhang, Phys. Rev. D47, 3065 (1993).

\item{[18]}{The infinite sphaleron energy in the true vacuum has nothing
to do with the generation of the baryon asymmetry. For a first order
phase transition, the baryon number is produced
 within and/or in front of the
bubble wall where $v \sim 0$, so
$E^{sphal} \sim 0$.}

\bye

\b

\bye